\documentclass[twocolumn,showpacs,prl,floatfix]{revtex4}

\ifx\pdfoutput\undefined
\usepackage{graphicx}
\else
\usepackage[pdftex]{graphicx}
\usepackage{epstopdf}
\fi

\usepackage{amsmath}
\usepackage[center]{subfigure}

\begin{document}
\title{Emergence of foams from the breakdown of the phase field crystal model} \author{Nicholas
  Guttenberg$^1$, Nigel Goldenfeld$^1$ and Jonathan Dantzig$^2$}
\affiliation{$^1$Department of Physics, 1110 West Green Street,
$^2$Department of Mechanical science and Engineering, 1206 West Green
Street,
  University of Illinois at Urbana-Champaign,
  Urbana, Illinois, 61801-3080.}

\begin{abstract}
The phase field crystal (PFC) model captures the elastic and
topological properties of crystals with a single scalar field at small
undercooling. At large undercooling, new foam-like behavior emerges. We
characterize this foam phase of the PFC equation and propose a modified
PFC equation that may be used for the simulation of foam dynamics. This
minimal model reproduces von Neumann's rule for two-dimensional dry
foams, and Lifshitz-Slyozov coarsening for wet foams.  We also measure
the coordination number distribution and find that its second moment is
larger than previously-reported experimental and theoretical studies of
soap froths, a finding that we attribute to the wetness of the foam
increasing with time.
\end{abstract}

\pacs{47.57.Bc, 64.70.D-, 47.54.-r}

\maketitle

Computational modeling of large-scale systems usually involves either
detailed molecular dynamics simulations, in which every particle must
be tracked, or a highly coarse-grained model in which the underlying
symmetries are known and are used to derive a continuum model from the
underlying physics. Molecular dynamics is limited to small systems
and/or to very short times, and coarse-grained models tend to fail at
places where the underlying symmetries are broken, such as at defects
and dislocations. The phase field crystal (PFC) model
\cite{elder2004modeling} is an intermediate approach with the advantage
of diffusive time scale, but with atomic scale resolution of molecular
dynamics. The PFC model can be used to capture the dynamics of defects
and dislocations in large crystals
\cite{elder2007phase,provatas2007using}, the dynamics of grain
interactions \cite{mellenthin2008phase}, molecular dynamics of
vacancies\cite{chan09molecular} and even nonlinear
elasticity\cite{CHAN09nonlinear_elasticity}.  In addition, it is
amenable to methods such as coarse-graining and adaptive mesh
refinement \cite{goldenfeld2005renormalization}.

The dimensionless phase field crystal equation \cite{elder2004modeling}
\begin{equation}\label{PFC.0}
\frac{\partial\psi}{\partial t} = \nabla^2 [ (\nabla^2+1)^2\psi + r \psi+\psi^3 ]
\end{equation}
is a density functional theory \cite{elder2007phase}, best thought of
as arising from phenomenological and symmetry considerations.  In
particular, this is the simplest class of models appropriate for
systems whose dynamics is governed by minimizing departures from
periodicity\cite{Brazovskii}, as opposed to the situation in other
materials processes, such as spinodal decomposition, where the dynamics
minimizes departures from spatial uniformity. In principle this
approach only holds for small values of the undercooling $\alpha \equiv
-r$, beyond which the strong nonlinearity may in principle overwhelm
the crystal's symmetries, leading to such artifacts as merger or
dissolution of the \lq atoms' of the model.

In this Rapid Communication, we  explore an interesting feature of the
phase field crystal model in the limit of large undercooling, which we
show turns out to facilitate a simple, scalar and minimal model of
foams. In the appropriate density regimes, the behavior of the phase
field crystal equation at large undercoolings is that the atoms of the
crystal lattice begin to merge. However, the interstitial spaces
between the atoms are preserved and become line solitons. The result is
a coarsening foam-like structure. We analyze the process by which this
instability in the equation of motion occurs, and use this
understanding to propose a minimal, modified PFC model that is capable
of describing quantitatively both wet and dry foams at the level of a
continuum, scalar theory that is computationally efficient.

{\it Equilibrium Phase Diagram:-} The phase diagram of the PFC model
has been computed for small undercoolings\cite{elder2004modeling}. We
shall use the same methods to construct the phase diagram at larger
values of the undercooling in order to see what may be found there.
First, however, we will convert the PFC equation and PFC energy to a
nondimensionalized form with respect to the equilibrium liquid
(constant phase) density. The energy minimizing density for the
constant phase is $\psi = \pm \sqrt{\alpha}$, so we will introduce a
nondimensional order parameter $\phi \equiv \psi/\sqrt{\alpha}$. This
gives us the following free energy:
\begin{equation}\label{PFC.1}
F = \int\limits_V \frac {1}{2} \phi (\nabla^2 + 1)^2 \phi + \alpha
\left(-\frac{1}{2} \phi^2 + \frac{1}{4} \phi^4 \right) dV
\end{equation}
and corresponding equation of motion:
\begin{equation}\label{PFC.2}
\frac{\partial\phi}{\partial t} = \nabla^2 ( (\nabla^2+1)^2 \phi + \alpha(-\phi + \phi^3))
\end{equation}

We then construct the phase diagram at large $\alpha$ by calculating
the energy minima of the one-mode approximations for the constant
phases $L_{\pm}: \phi = {\phi_0}$, where the subscript $\pm$ refers to
the sign of the average density $\phi_0$; the striped phase $S: \phi =
\phi_0 + A_S \cos(x)$, where $A_S = \sqrt{\alpha(4-3\phi_0^2)/3}$; and
the two triangular lattices $\Delta_{\pm}: \phi = \phi_0 + (A_\Delta
\pm B_\Delta) ( \cos (\vec{k_1} \cdot \vec{r}) + \cos (\vec{k_2} \cdot
\vec{r}) + \cos(\vec{k_3} \cdot \vec{r}))$, where the $\vec{k_j}$ are
the lattice vectors of the regular triangular lattice, $A_\Delta =
-\alpha \phi_0/5$ and $B_\Delta = \sqrt{\alpha(15-36\phi_0^2)}/15$.

Even though the one-mode approximation is not accurate at these large
values of $\alpha$, we use this approximation to understand
heuristically the behavior observed, thus motivating our form for the
modified phase field crystal model introduced below. Our main results,
for the modified phase field crystal model, are fully time-dependent
and independent of this approximation.

\begin{figure}[htb]
\includegraphics[width=0.9\columnwidth,angle=0]{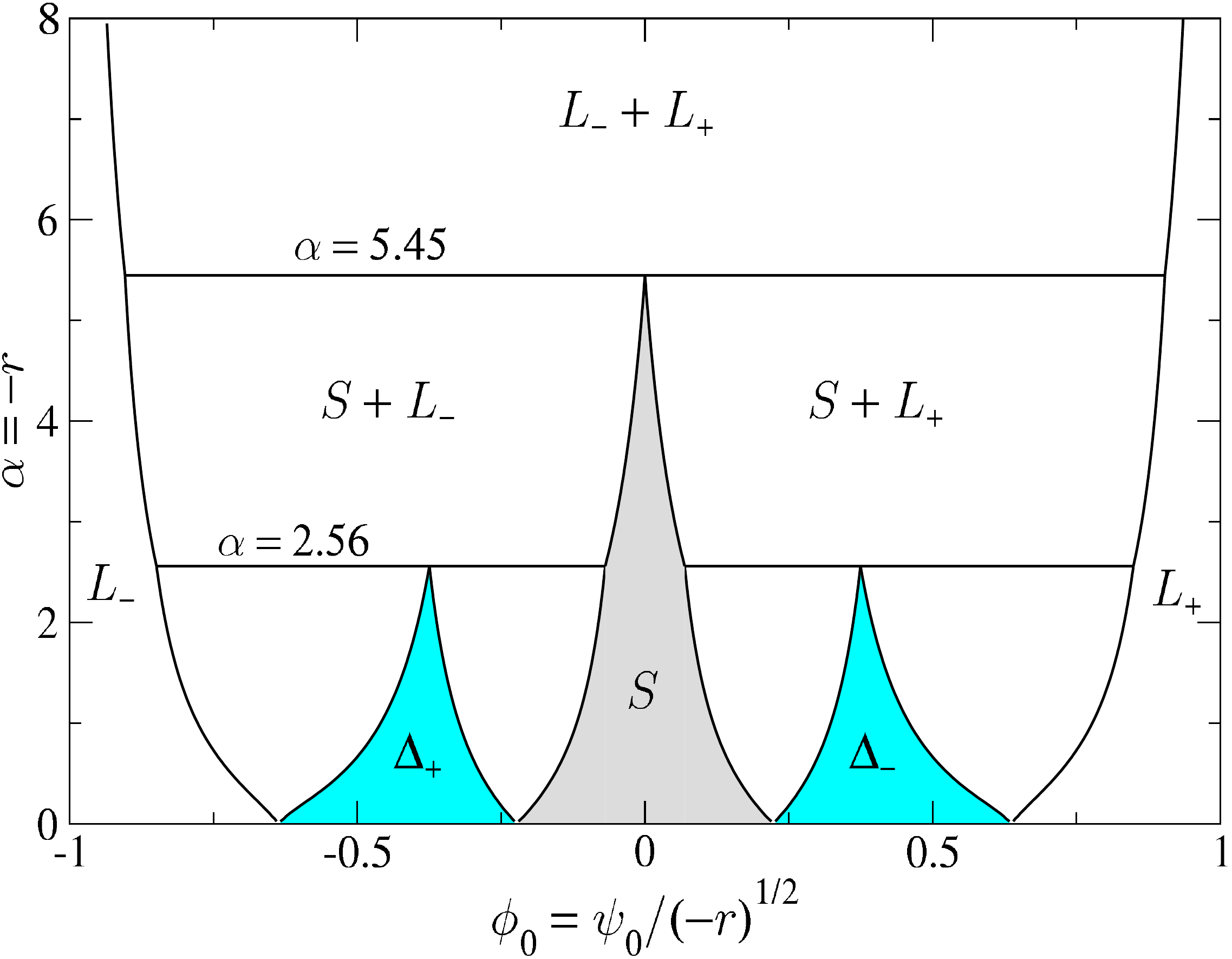}
\caption{(Color online). Phase diagram associated with the free energy in
  Eq.~(\ref{PFC.1}). Invariant reactions occur at
$\alpha=2.56$ and
$\alpha=5.45$. No further topological changes occur for $\alpha > 8$.}
\label{Fig1}
\end{figure}

The equilibrium phase diagram is obtained by substituting the various
ansatz into Eqn.~(\ref{PFC.2}) and performing a common tangent
construction. The result is shown in Fig~\ref{Fig1}. We find a
series of previously undiscovered invariant points as $\alpha$
increases. At $\alpha=2.56$, the coexistence region between the
striped and triangular phases disappears, giving rise to a coexistence
between stripes and one of the liquid phases. Then, at $\alpha=5.45$
the striped phase vanishes, leaving only a region of immiscible
liquid-liquid coexistence, and the regions comprised entirely of the
liquid phases.

This can be understood by considering the influence of the two terms
in the free energy. At small $\alpha$, the wavelength selection term
that was added to construct the PFC equation is dominant. As $\alpha$
increases, the local term that drives $\phi$ to have the values of
$\pm 1$ becomes dominant, so that eventually the wavelength selection
term $(1 + \nabla^2)^2)\phi$ is a small perturbation and the
equilibrium phases are determined by the average value of
$\phi_0^2/2-\phi_0^4/4$, which is minimized by the constant phases.

{\it Dynamics:-} The phase  diagram we  have just computed  represents
the  behavior of this  system in  its final  equilibrium state.
However,  the approach toward that equilibrium changes as $\alpha$
increases. Starting from a hexagonal lattice,  the system attempts  to
coarsen into  two regions, one composed of  $L_+$, the other composed
of  $L_-$. However, because of the small  residual wavelength
selection, there is  a finite energy cost to  removing spatially
patterned structures. The  consequence of this  is  that  the  PFC
`atoms'  are  dynamically  conserved  by  this infinitesimal energy
barrier, which halts the coarsening dynamics. If the average density of
the  system is shifted, e.g., by deliberately reducing $\phi_0$ over
time, this can provide enough energy to destabilize  the PFC atoms, but
stripe-like  cell walls will still remain stable for a longer period as
they can move perpendicular to their length to create large bubbles of
one of the liquids.

\begin{figure}[htb]
  \includegraphics[width=\columnwidth,angle=0]{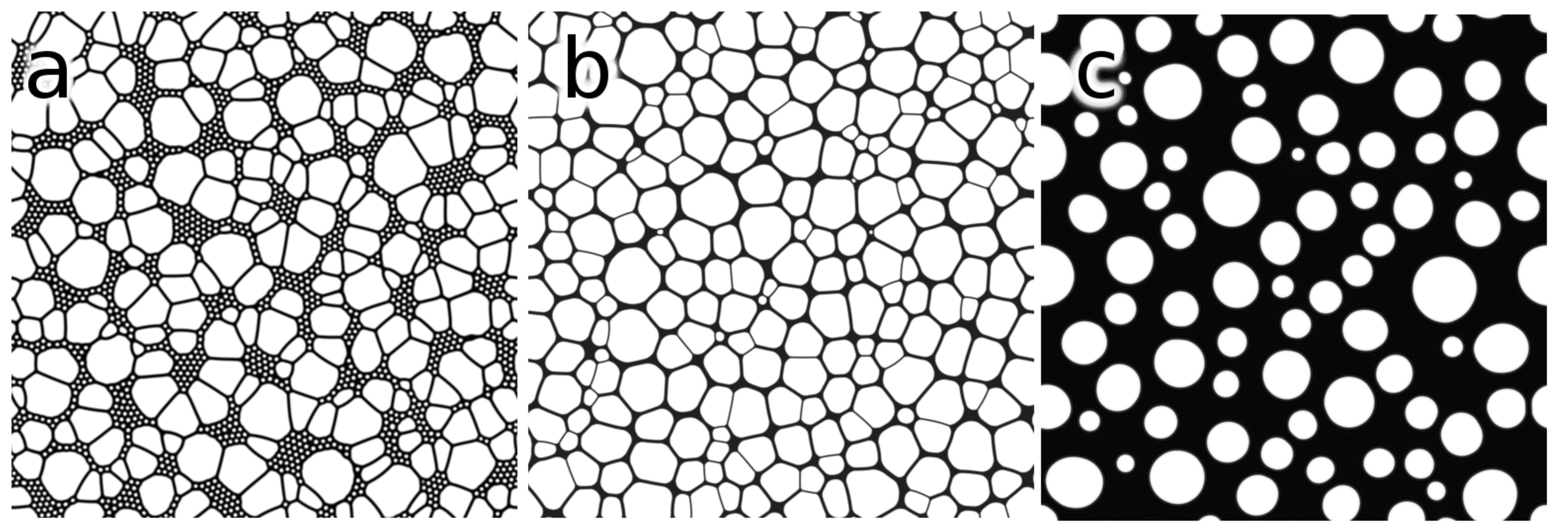}
  \caption{a. Final state obtained from Eqn.~(\ref{PFC.2}) with
    $\alpha=16$ and $\phi_0=-0.41$ after coarsening. Residual atoms
    coexist with large foam cell regions. b. Simulation using the
    modified PFC equation for $\alpha=20$, quenched with $\phi_0=0.4$
    initially, and then drained to $\phi_0=-0.6$. The result is a dry
    foam structure with no residual atoms. c. Simulation of the
    modified PFC equation for $\alpha=20$ and $\phi_0=0.25$, allowed
    to coarsen without draining. The result is
    a wet foam structure with circular bubbles. } \label{Fig2}
\end{figure}

The end result of this is that a foam-like state appears
(Fig~\ref{Fig2}a), which coarsens to a stationary point at which
there are no more free adjustments that can be made to approach the
equilibrium state. Thus, even a small perturbative addition of
wavelength selection is sufficient to stabilize a foam-like structure.

Although this behavior is interesting from the point of view of
understanding the properties of the phase field crystal model, it does
not behave like a physical foam. The arrested coarsening and inability
to get rid of residual atoms prevent this from being used as a model
for studying coarsening foams. We now alter the free energy so as to
destabilize the residual atoms while retaining the overall cellular
structure, so that we may recover physical foam dynamics.

In effect what we wish to do is to weaken the wavelength selection
while encouraging $k=0$ structures. A stripe has one $k=0$ direction
and one direction with the selected wavelength, whereas an atom has
two directions with the selected wavelength. If the selected
wavelength sits at a local energy minimum, but the global minimum is
at $k=0$, then this should help remove residual atoms in favor of
bubble interfaces.

The PFC wavelength selection operator $(\nabla^2+1)^2$ corresponds in
$k$-space to
\begin{equation}
E(k) = (1-k^2)^2
\end{equation}
We modify this by introducing two parameters $k_0$ and $b$
\begin{equation}
E_{mod}(k) = (k/k_0)^2(1-(k/k_0)^2)^2+b (k/k_0)^2
\end{equation}
and then choose $k_0$ so as to retain minima at $k=\pm1$: $k_0=[3/(2+
\sqrt{1-3b})]^{1/2}$. The modified form of the free energy in real space becomes
\begin{align}\label{PFC.mod}
F_{mod} = \int\limits_V &\frac {1}{2} \phi \left({1 \over k_0^2} \nabla^2
\left({1 \over k_0^2}\nabla^2 + 1\right)^2 -{b \over k_0^2}\nabla^2 \right) \phi
\nonumber \\
& + \alpha \left(-\frac{1}{2} \phi^2 + \frac{1}{4} \phi^4 \right) dV
\end{align}
We can then vary $b$ to control the relative depths of the minima at
$k=\pm 1$. The most extreme effect is achieved at $b=-1/3$, at which
the minima at $k=\pm 1$ disappear. We use this value of $b$
throughout the rest of this paper.

The dynamics of the modified PFC equation give rise to the structures
shown in Figs.~\ref{Fig2}b and \ref{Fig2}c. If the system is
quenched from a disordered state, one recovers Lifshitz-Slyozov
coarsening dynamics \cite{bray2002theory} corresponding to the physical
behavior of a very wet foam. On the other hand, if the system is
drained from a state with positive average density to one with
negative average density, polygonal cell walls form and coarsen by
topological rearrangements.

{\it Results:-} We measured the coarsening dynamics and statistics of
the eventual scaling state of the modified PFC equation in order to
compare with physical foams. In order to measure the coarsening we are
interested in the quantity $\langle r \rangle(t)$, the average bubble
radius. We may easily count the total number of bubbles and so
determine the average bubble area as $\langle A \rangle (t) =
A_{total}/N(t)$. If we assume that deviations from circular geometry
are small, then we may approximate $\langle r \rangle (t) \approx
\sqrt{\langle A
  \rangle(t)}$.

\begin{figure}[htb]
\includegraphics[width=\columnwidth,angle=0]{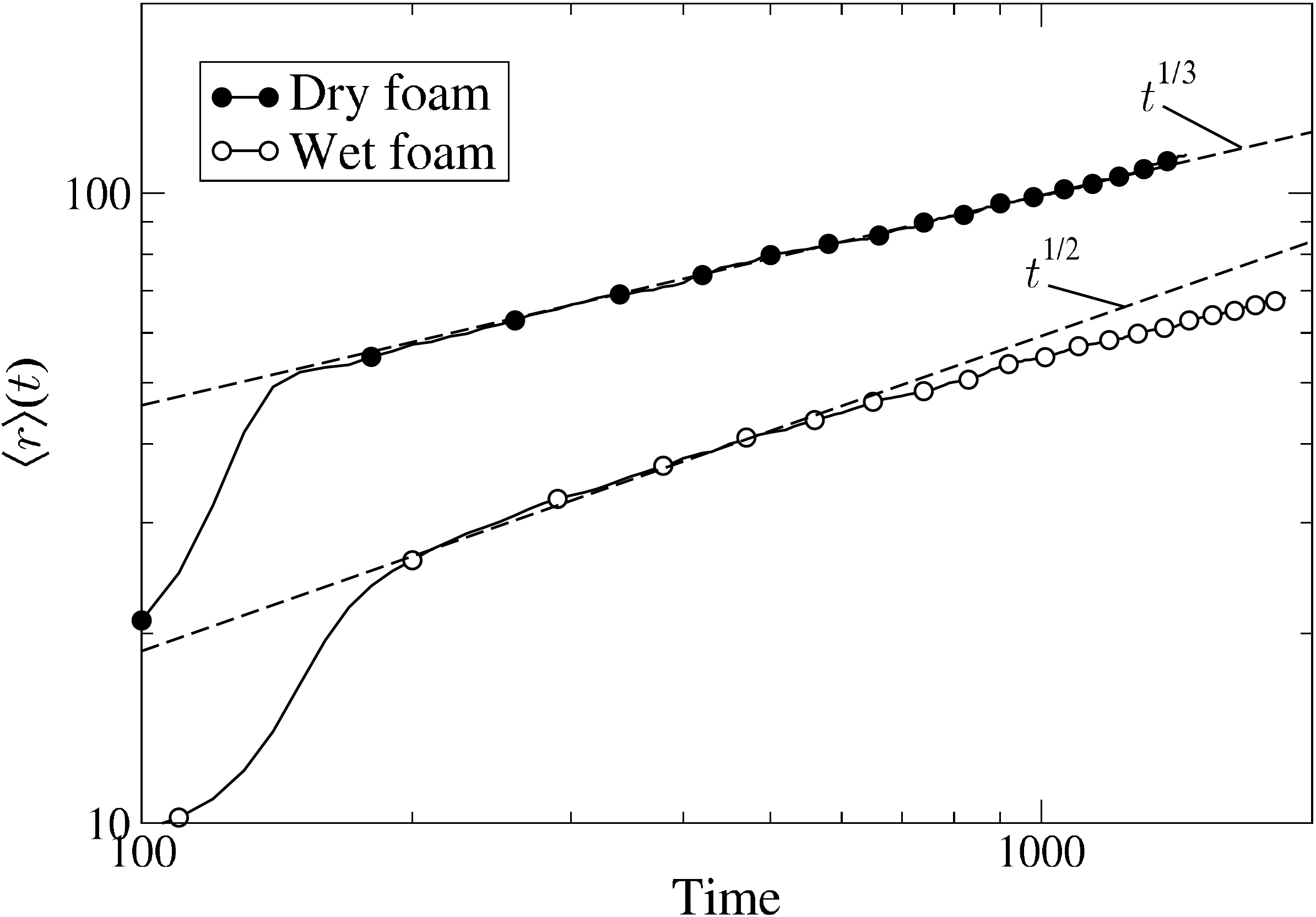}
\caption{Coarsening behavior of wet and dry foams in the modified PFC
equation with $\alpha=20$, compared with the corresponding theoretical
coarsening laws.} \label{Fig3}
\end{figure}

For a wet foam, we expect the bubble growth to proceed as diffusive
grain growth, so that $\langle r \rangle (t) \propto t^{1/3}$. As an example, we
simulate a 1024x1024 system with $\alpha=20$ and quench into an
average density $\phi_0=0.25$. We observe a scaling $\langle r \rangle
\propto t^{0.34}$ (Fig \ref{Fig3}, filled circles), consistent with the
predicted growth law.

For a dry foam in two dimensions, von Neumann's law for bubble area
growth \cite{von1952metal} is
\begin{equation}
\frac{dA}{dt} = \kappa (n-6)
\end{equation}
where $\kappa$ is an effective diffusivity, and $n$ is the
coordination number of the bubble. This implies that the average
bubble radius should scale as $\langle r \rangle(t) \propto t^{1/2}$.
To measure the coarsening of a dry foam, we quench from an average
density $\phi_0=0.3$ and drain slowly to a density of $\phi_0=-0.4$.
We start measuring the coarsening dynamics after we have stopped
draining. However, our reference point for $t=0$ is still the point of
the quench. We observe a scaling $r \propto t^{0.43}$, significantly
slower than the predictions of von Neumann's law (Fig \ref{Fig3}, open
circles). At late times in the simulation, the bubble interfaces start
to become wet. If we restrict our analysis to times shortly after
coarsening begins, the resulting fit exponent increases to $r\propto
t^{0.47}$. This indicates that we are probably seeing the effect of
not having a fully dry foam at any point in time.

As two-dimensional froths coarsen, they are expected to reach a
self-similar scaling state in which the normalized moments of the
distribution of areas and of coordination number are expected to become
time independent. The second moment of the coordination number
distribution has been used as a probe of this scaling state. Glazier
and Weaire \cite{glazier1992kinetics} predict that the coordination
number distribution should eventually reach a universal limiting
scaling state with a second moment $\mu_2 = 1.4$, with a strongly
non-monotonic transient behavior.

\begin{figure}[htb]
\includegraphics[width=0.9\columnwidth]{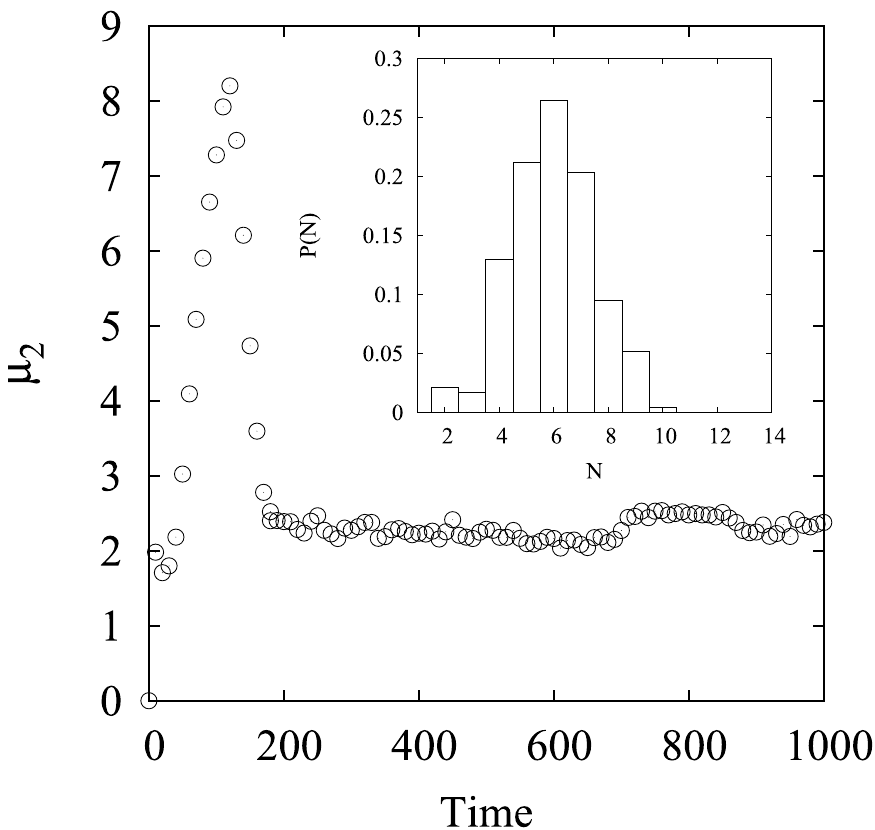}
  \caption{Evolution of the second moment of the bubble coordination
    number in a dry foam realized by the modified PFC equation. The
    bubble distribution broadens during the transition to power-law
    coarsening, after which it reduces to a steady value of
    $2.31\pm0.01$. The inset shows a particular distribution at
    $t=1800$. } \label{Fig4}
\end{figure}

We measure the coordination number distribution of the dry foam by a
watershed algorithm. \cite{watershed} We superimpose a grid over the
system and fill each bubble with a unique integer. We then allow these
integers to propagate to neighboring unoccupied cells, and then find
the total number of different integers that a given bubble comes into
contact with. We observe a similar time-dependence of the second moment
of our distribution, but the measured limiting value $\mu_2 = 2.31 \pm
0.01$ is significantly different from 1.4 (see Fig \ref{Fig4}). In
other models and simulations of coarsening 2D foams, the observed
limiting value of $\mu_2$ has been variously reported as $1.9$
\cite{marder1987soap}, and $1.7$ \cite{flyvbjerg1993model}, and $1.5$
\cite{glazier1990coarsening}, and experiments have measured from $1.4$
\cite{stavans1989soap} in soap froths to values as small as $0.14$ and
$0.22$ in magnetic froths \cite{elias1997two}.

We conclude that the
second moment is either not a universal quantity, or else that there
are strong transients which make any universal scaling regime difficult
to observe in practice.  We note that our foam is not completely dry,
and  that the absence of a drainage mechanism implies that the foams
becomes wetter the more they coarsen.  Whether real foams undergo a
corresponding change of regime is not clear.

We have found that in the limit of large undercooling, the phase field
crystal equation produces topologically stabilized foams. These foams
are a consequence of the residual wavelength selection, which acts as
a singular perturbation that prevents the destruction of cell walls
and forces the coarsening to take place via topological
rearrangements. The fact that this behavior emerges from a minimal
model such as the phase field crystal equation suggests a general
mechanism by which foams may occur in natural systems.

The ingredients of the phase field crystal equation are a driving
force towards certain equilibrium densities, some sort of spatial
relaxation (diffusion, viscosity, and the like), and some competing
source of wavelength selection. Such a system, in the limit where the
wavelength selection is weak compared to the other forces, would
likely give rise to a foam. This mathematical structure can be seen in
models of magnetic froths \cite{elias1997two}, Type-I
superconductors \cite{prozorov2008suprafroth}, and in models of
polygonal cells found in melting snow \cite{tiedje2006radiation}.

However, the foams produced in this limit of the PFC equation have
unphysical properties. The wavelength selection also preserves atoms:
bubbles whose diameter is comparable to the width of the cell
walls. This can be ameliorated by modifying the PFC free energy to
penalizes atoms while encouraging stripes. As a result, the
coarsening dynamics of real foams are recovered although the
distribution of bubbles in the resultant scaling state appears to be
somewhat different.

If these differences are understood, the modified PFC equation may be
a useful tool in modelling foams, as it is simpler than other existing
methods such as Q-Potts models \cite{jiang1996extended} and minimal
surface evolution \cite{phelan1995computation}, and is fairly easy to
simulate, having only one field and a spatial structure that is easily
treated with spectral methods.

We are grateful to B.~Athreya for discussions about the stability
of the PFC equation. N.~Guttenberg was supported by a University of
Illinois Distinguished Fellowship, and J. Dantzig acknowledges the
support of the US Dept. of Energy, under subcontract 4000076535.

\bibliographystyle{apsrev}
\bibliography{pfcfoam}

\end{document}